\def\year{2018}\relax
\documentclass[letterpaper]{article}
\usepackage{aaai18}
\usepackage{times}
\usepackage{helvet}
\usepackage{courier}

\usepackage[utf8]{inputenc}
\usepackage[T2A,LAE,T1]{fontenc}
\usepackage[russian,arabic,USenglish]{babel}

\usepackage{amsmath}
\usepackage{bbm}
\usepackage{balance}       
\usepackage{graphics}      

\usepackage{txfonts}
\usepackage{color}
\usepackage{booktabs}
\usepackage{textcomp}
\usepackage{graphicx}
\usepackage{caption}
\usepackage{subcaption}
\usepackage{balance}  
\usepackage{blindtext}
\usepackage{url}

\usepackage{xargs} 
\usepackage[colorinlistoftodos,prependcaption,textsize=tiny]{todonotes}
\newcommandx{\ak}[2][1=]{\todo[linecolor=red,backgroundcolor=red!25,bordercolor=red,#1]{#2}}
\newcommandx{\vicenc}[2][1=]{\todo[linecolor=blue,backgroundcolor=blue!25,bordercolor=blue,#1]{#2}}

\begin{document}
\title{Online Petitioning Through Data Exploration and What We Found There:\\ A Dataset of Petitions from Avaaz.org\footnote{This version extends the paper presented at \mbox{ICWSW-18} with an appendix including the reasons, provided by \textit{Avaaz.org}, about the anomalies detected when exploring the dataset.}}

\author{Pablo Aragón\textsuperscript{1,2}~ Diego Sáez-Trumper\textsuperscript{1}~ Miriam Redi\textsuperscript{3}~\\ {\bf \Large Scott A. Hale\textsuperscript{4}~ Vicenç Gómez\textsuperscript{1}~ Andreas Kaltenbrunner\textsuperscript{1}}\\\\
\textsuperscript{1}Universitat Pompeu Fabra~~~~\textsuperscript{2}Eurecat, Centre Tecnològic de Catalunya\\\textsuperscript{3}King's College, London~~~~\textsuperscript{4}Oxford Internet Institute, University of Oxford\\~\\}

\let\ACMmaketitle=\maketitle
\renewcommand{\maketitle}{\begingroup\let\footnote=\thanks \ACMmaketitle\endgroup}

\maketitle
\begin{abstract}
The Internet has become a fundamental resource for activism as it facilitates political mobilization at a global scale. Petition platforms are a clear example of how thousands of people around the world can contribute to social change. \textit{Avaaz.org}, with a presence in over 200 countries, is one of the most popular of this type. However, little research has focused on this platform, probably due to a lack of available data.

In this work we retrieved more than 350K petitions, standardized their field values, and added new information using language detection and named-entity recognition. To motivate future research with this unique repository of global protest, we present a first exploration of the dataset. In particular, we examine how social media campaigning is related to the success of petitions, as well as some geographic and linguistic findings about the worldwide community of \textit{Avaaz.org}. We conclude with example research questions that could be addressed with our dataset.
\end{abstract}

\section{Introduction}

Petition signing has long been one of the most popular political activities with a history extending back to at least the Middle Ages~\cite{fox2012}. After a decline in the 20th century, petitioning has achieved new prominence through online, ``e-'' petition platforms. Online petitions are often disseminated on social media, and their low-costs, low-barriers to entry may bring new people into the political process~\cite{margetts2015political}. 

Despite the popularity of petition platforms, this computer-mediated form of civic participation has also come under criticism. A review of online petitioning over 10 years in Europe and the United Kingdom concluded that, although these experiences were mostly positive, there was no solid evidence about significant impact~\cite{Panagiotopoulos2012}. Indeed, petition platforms are seen as the essence of the so-called `slacktivism' or `clicktivism', where the main effect is not to impact real life but to enhance the feel-good factor for participants~\cite{christensen2011slacktivism}. This criticism has fueled a skeptical view of activism through social media~\cite{gladwell2010small} and of the promise that the Internet would `set us free'~\cite{Morozov:2011:NDD:1964879}. Nevertheless, it cannot be ignored that online petitioning has a proven ability to originate public policies of great impact, e.g., the \textit{Unlocking Consumer Choice and Wireless Competition Act}~\cite{uscongress}.

In the last decade, much research has focused on analyzing data from government platforms in countries like Germany~\cite{jungherr2010,lindner2011}, the United Kingdom~\cite{wright2012,hale2013,wright2015}, and the United States~\cite{dumas2015,margetts2015political,yasseri2017}. As these platforms belong to public institutions, these studies are of interest because of the potential of petitions to influence policy making. However, institutional platforms are restricted by definition to specific territories, and do not take advantage of the global scope offered by the Internet.

At the global level, \textit{Change.org} and \textit{Avaaz.org} are two paradigmatic examples of online petition platforms able to engage millions of people around the world. Nevertheless, there are few empirical studies of these communities. For \textit{Change.org}, launched in 2007 by a for-profit corporation, research has analyzed user behavior~\cite{huang2015}, success factors~\cite{7427430} and  gender patterns~\cite{peixoto}. These studies relied on data from the API but, since October 2017, it is no longer supported\footnote{\url{https://help.change.org/s/article/Change-org-API}}. For \textit{Avaaz.org}, also launched in 2007 but founded by non-profit organizations, the only case study to date focused on whether this community fulfills certain basic democratic dimensions~\cite{horstink2017online}. This study reported serious problems of
transparency and accountability because of the lack of information about their activities. In fact, \textit{Avaaz.org} does not provide an open data API, which is a major barrier to research in this field.

To overcome the lack of available data from global petition platforms, we present in this article an open dataset of petitions from \textit{Avaaz.org}\footnote{Available at \url{https://dataverse.mpi-sws.org/dataverse/icwsm18}}. In the following section we describe how we obtained the data in line with technical and legal requirements, and the structure of the dataset. We then explore the data to provide some findings of interest. To motivate future work, we conclude by offering example research questions that could be addressed with our dataset. 

\section{Data Collection}

\begin{figure}[h]
 \centering
   \includegraphics[clip, trim=14cm 0cm 14cm 0cm, width=0.765\linewidth]{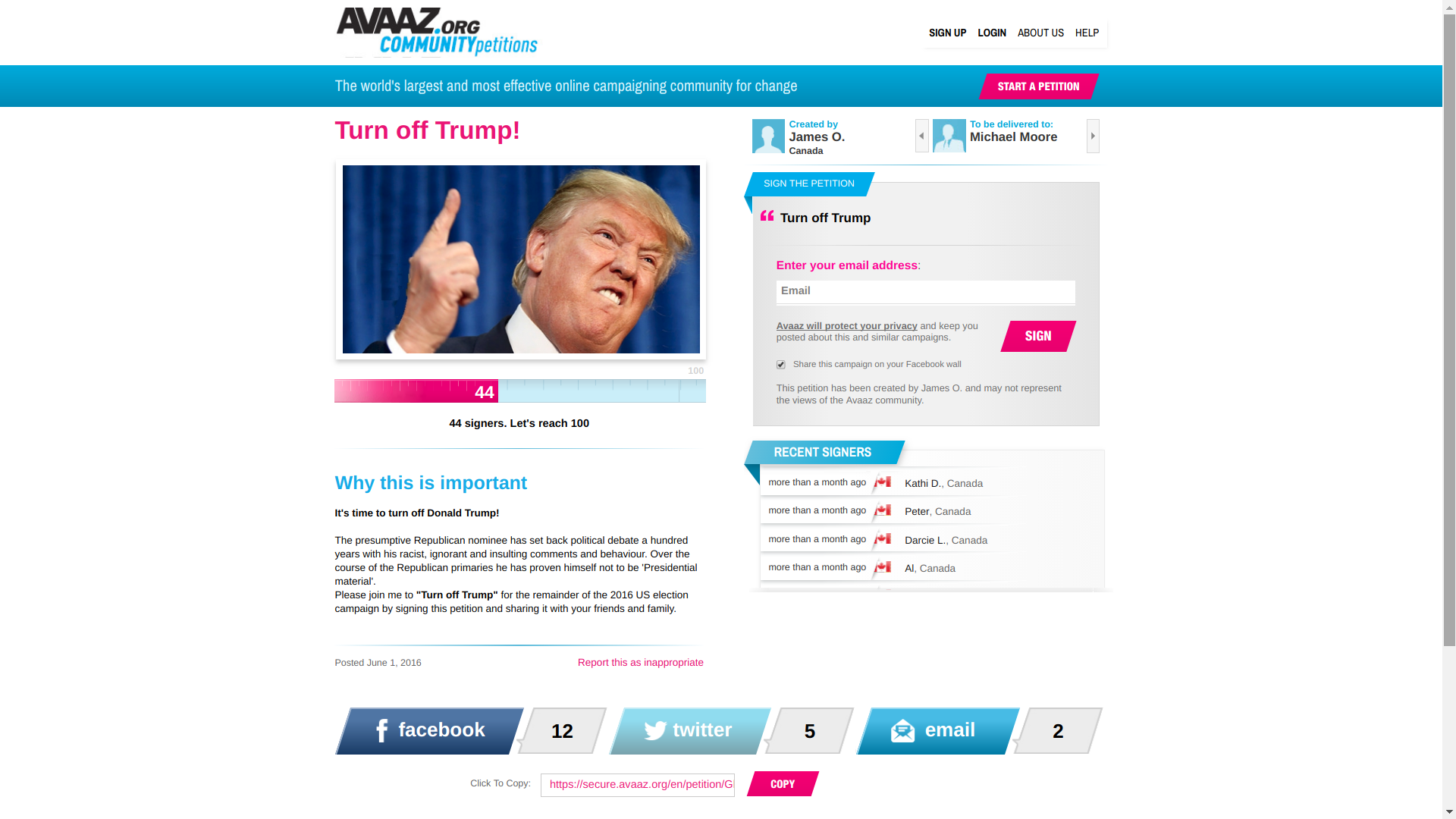}
 \caption{Web page of a petition in \textit{Avaaz.org} \url{https://secure.avaaz.org/en/petition/Global_media_consumers_Turn_off_Trump/}.}
\label{fig:trump}
\end{figure}

To generate the dataset of petitions from \textit{Avaaz.org}, we implemented a web crawler based on the incremental nature of their numerical ids. First, for a given petition id, a script sent a request to the AJAX endpoint of \textit{Avaaz.org} and retrieved the corresponding URL. Then, with the petition URL, another script fetched and parsed the HTML to extract and store the corresponding metadata (an example petition page is shown in Figure~\ref{fig:trump}). The crawling process was done in August 2016 and the petition ids ranged from 1 to 382979 (which was the latest petition at that time). After excluding deleted pages, we obtained a dataset of 366,214 petitions.

It is important to highlight two issues taken into account when the crawler was designed. First, the machine-readable robots.txt file on \textit{Avaaz.org} does not specify any restrictions\footnote{\url{https://secure.avaaz.org/robots.txt}}. Second, every page fetched by the crawler specified a \textit{Creative Commons Attribution 3.0 Unported License} in the footnote. Therefore, our dataset is released under the same terms.

\subsection{Structure of the dataset}

The metadata of the online petitions were processed to produce a standardized and enriched dataset. Besides the id and URL, each petition contains the following fields:

\begin{itemize}
\item \texttt{title} (string): Title of the petition (limited to 100 characters). Following the official guidelines about how to write a petition title~\cite{avaaz2018}, many of them include the person, organization and/or location it addresses.

\item \texttt{description} (string): Description of the petition. 

\item \texttt{author} (string): Name of the user who authored the petition. To preserve anonymity, \textit{Avaaz.org} includes only the given name and the first initial of the family name.

\item \texttt{date} (timestamp): Date when the petition was published, from December 2011 to August 2016. Because dates were originally found in different languages including different writing systems (Latin, Arabic, Cyrillic, Kana, Hebrew, and Greek), we standardized them as \textit{yyyy-MM-dd}.

\item \texttt{country\_{}name} (string): Name of the country of origin of the author. Users are able to customize this field in their profiles. Otherwise, the value is obtained by \textit{Avaaz.org} directly from the user's IP address.
\item \texttt{country\_{}code} (string): As well as \texttt{date}, country names were originally found in different languages and writing systems, e.g., Turkey was also found as  \foreignlanguage{arabic}{تركي}, Türkiye, Türkei, Turchia, Turquie, Turquía, and \foreignlanguage{russian}{Турция}. Therefore, we standardized countries using ISO 3166-1 alpha-3 codes\footnote{\url{https://unstats.un.org/unsd/methodology/m49/}}.
\item \texttt{sign} (integer): Number of signatures at the time the petition was crawled. 

\item \texttt{target} (integer): Number of signatures set as a goal by the platform/creator (this value may change as petitions receive signatures).

\item \texttt{ratio} (float): Ratio between the two proceeding fields.
\item \texttt{facebook\_{}count} (integer): Number of shares on Facebook.
\item \texttt{twitter\_{}count} (integer): Number of shares on Twitter.
\item \texttt{whatsapp\_{}count} (integer): Number of shares on WhatsApp.
\item \texttt{email\_{}count} (integer): Number of shares via email.

\item \texttt{lang\_{}code} (string): Language detected on the concatenation of title and description using a plugin for Apache Nutch project\footnote{\url{https://wiki.apache.org/nutch/LanguageIdentifier}} (based on n-grams of over 50 languages). The resulting language is standardized with a ISO-639 2-letter code\footnote{\url{http://www.loc.gov/standards/iso639-2/php/code_list.php}}.

\item \texttt{lang\_{}prob} (float): Probability of success given by the language detection plugin.

\item \texttt{people} (multivalued string): The names of people found within the concatenation of the title and description fields using the Stanford Named Entity Recognizer (NER)~\cite{Finkel2005Incorporating}. Results are only provided when the detected language was English, Spanish or German (available languages).

\item \texttt{organizations} (multivalued string): The names of organizations found using the NER, as done for \texttt{people}. 

\item \texttt{locations} (multivalued string): Locations found using the NER, as done for \texttt{people}. 

\item \texttt{miscellany} (multivalued string): Other entities found using the NER, as done for \texttt{people}. 
\end{itemize}

\begin{figure*}
\centering
\begin{subfigure}[b]{0.24\linewidth}
   \includegraphics[clip, trim=0cm 0cm 0cm 0cm,width=1\linewidth]{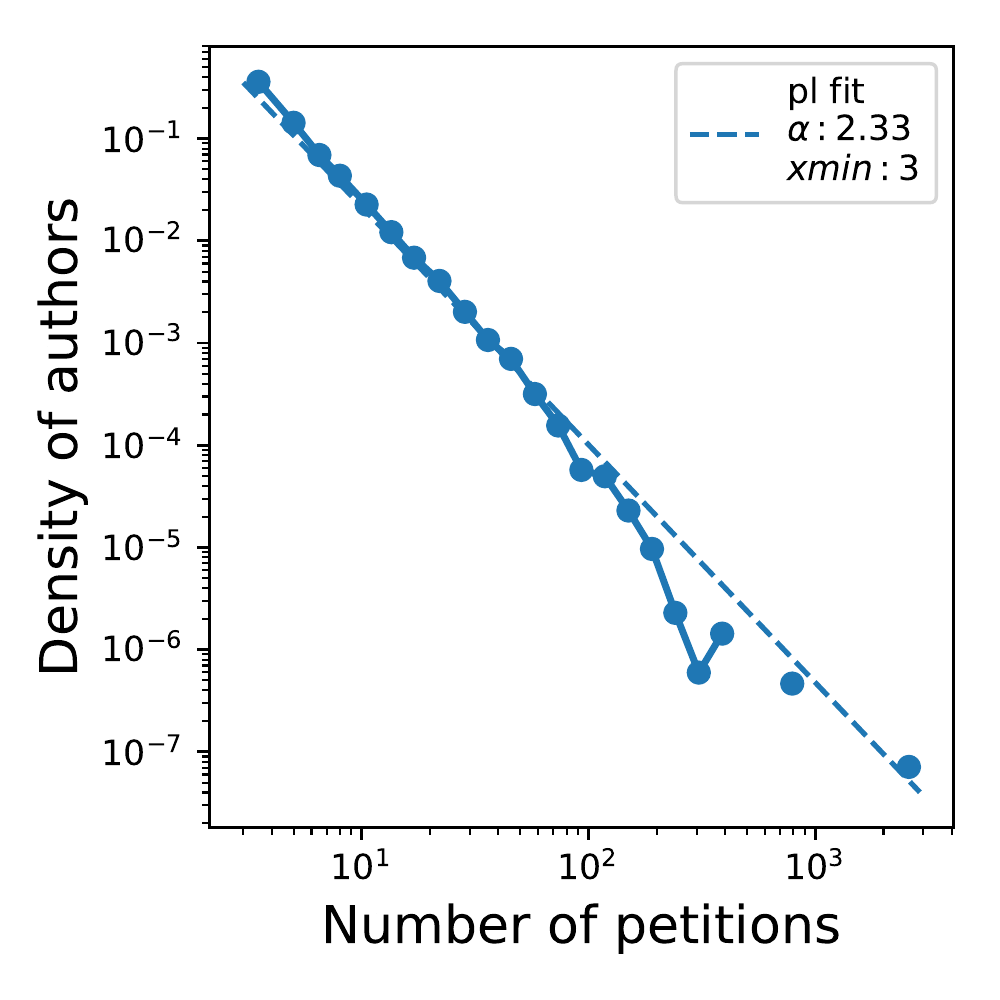}
   \caption{~}
   \label{fig:overview_authors} 
\end{subfigure}
\begin{subfigure}[b]{0.24\linewidth}
   \includegraphics[clip, trim=0cm 0cm 0cm 0cm,width=1\linewidth]{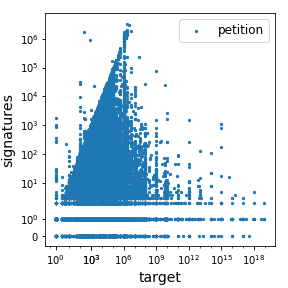}
   \caption{~}
   \label{fig:overview_scatter} 
\end{subfigure}
\begin{subfigure}[b]{0.24\linewidth}
   \includegraphics[clip, trim=0cm 0cm 0cm 0cm,width=1\linewidth]{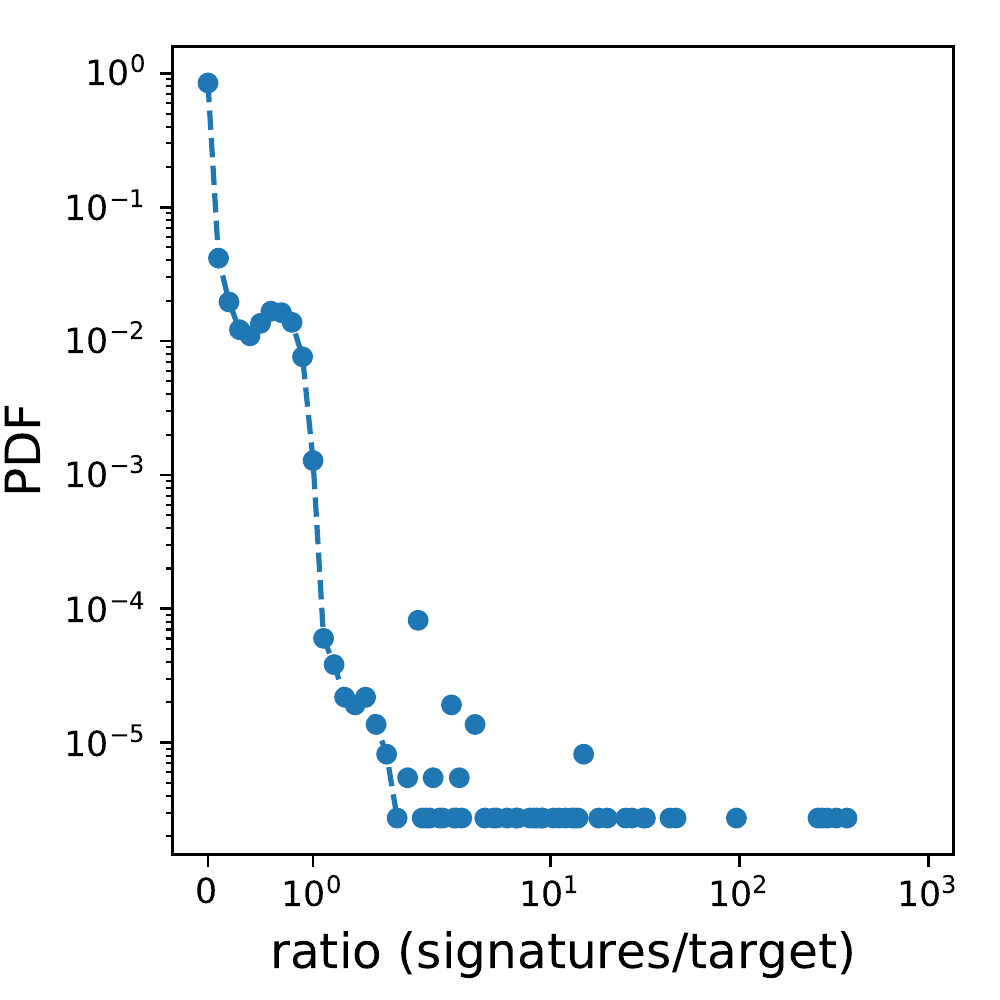}
   \caption{~}
   \label{fig:overview_ratio} 
\end{subfigure}
\begin{subfigure}[b]{0.24\linewidth}
   \includegraphics[clip, trim=0cm 0cm 0cm 0cm,width=1\linewidth]{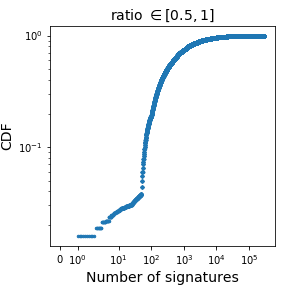}
   \caption{~}
   \label{fig:overview_ratio051} 
\end{subfigure}
\caption{General descriptive plots of the dataset of petitions: a) Distribution of authors by the number of petitions; \mbox{b) Target vs signatures;} c) Distribution of petitions by ratio; d) Cumulative distribution of petitions by signatures for petitions with a ratio $\in [0.5, 1]$.}
\label{fig:overview}
\end{figure*}

\section{Data Exploration}
In this section we explore the petitions in our dataset to illustrate its content and relevance for research. We first provide a general overview regarding authors and signatures, and then inspect the link between signatures and shares on social media platforms. Finally, we examine some geographical and multilingual findings about the worldwide community of \textit{Avaaz.org}.

\subsection{General overview}

Figure~\ref{fig:overview} presents general descriptive statistics of petitions in relation to authors and signatures. The plot in Figure~\ref{fig:overview_authors} shows the distribution of authors by the number of petitions. As expected, the distribution appears to follow a power law: most authors only published a few petitions. However, there is a small group of authors with over 1,000 petitions each. We examined this group and found that the most prolific user (2,895 petitions) is named \textit{selenium s.}, located in Afghanistan and the United States, and authored petitions like \textit{``TEST - AUTOMATED - Who: TEST - AUTOMATED - What''} or \textit{``selenium1440423202: selenium1440423202''}. This might indicate that these petitions were automatically generated with Selenium\footnote{\url{http://www.seleniumhq.org}}, a popular web browser automation tool\footnote{An extended explanation of this user, provided by \textit{Avaaz.org} after submitting the camera-ready copy of this article, is described in the appendix.}. Although much recent research has been devoted to  characterize bots in social media platforms~\cite{Chu:2010:TTH:1920261.1920265,Truthy_icwsm2011class,Ferrara:2016:RSB:2963119.2818717}, most studies have focused on Twitter. To the best of our knowledge, this is the first evidence of bots operating in an online petition platform.

To assess the success of petitions in our dataset, we present a scatter plot of petitions' target number of signatures compared to their actual number of signatures (Figure~\ref{fig:overview_scatter}). These dimensions are not correlated and the plot reveals that the number of signatures rarely exceeds the targets. This finding is explicit in Figure~\ref{fig:overview_ratio} which shows the distribution of petitions by ratio, i.e., the number of signatures for each petition divided its target. We should note that, although the fraction of petitions decreases as the ratio increases, this is not the case for petitions with \mbox{ratio $\in [0.5, 1]$}. The observed peak could be the result of targets automatically updating when the number of signatures reaches a specific threshold. This is a relevant design feature in some online petition platforms that could encourage more signatures by magnifying the importance of a new signature to reach the target~\cite{margetts2012social,frey2004social}. For this reason, we present in Figure~\ref{fig:overview_ratio051} the cumulative distribution of petitions by their number of signatures for petitions with ratio $\in [0.5, 1]$. The plot shows two trends: one from 1 to 99 signatures and the other above 100 signatures, which is the default initial target on the platform. It appears the target number of signatures is mostly likely to automatically update after a petition has at least 100 signatures.

\begin{figure*}
\centering
\begin{subfigure}[b]{0.24\linewidth}
   \includegraphics[clip, trim=0cm 0cm 0cm 0cm,width=1\linewidth]{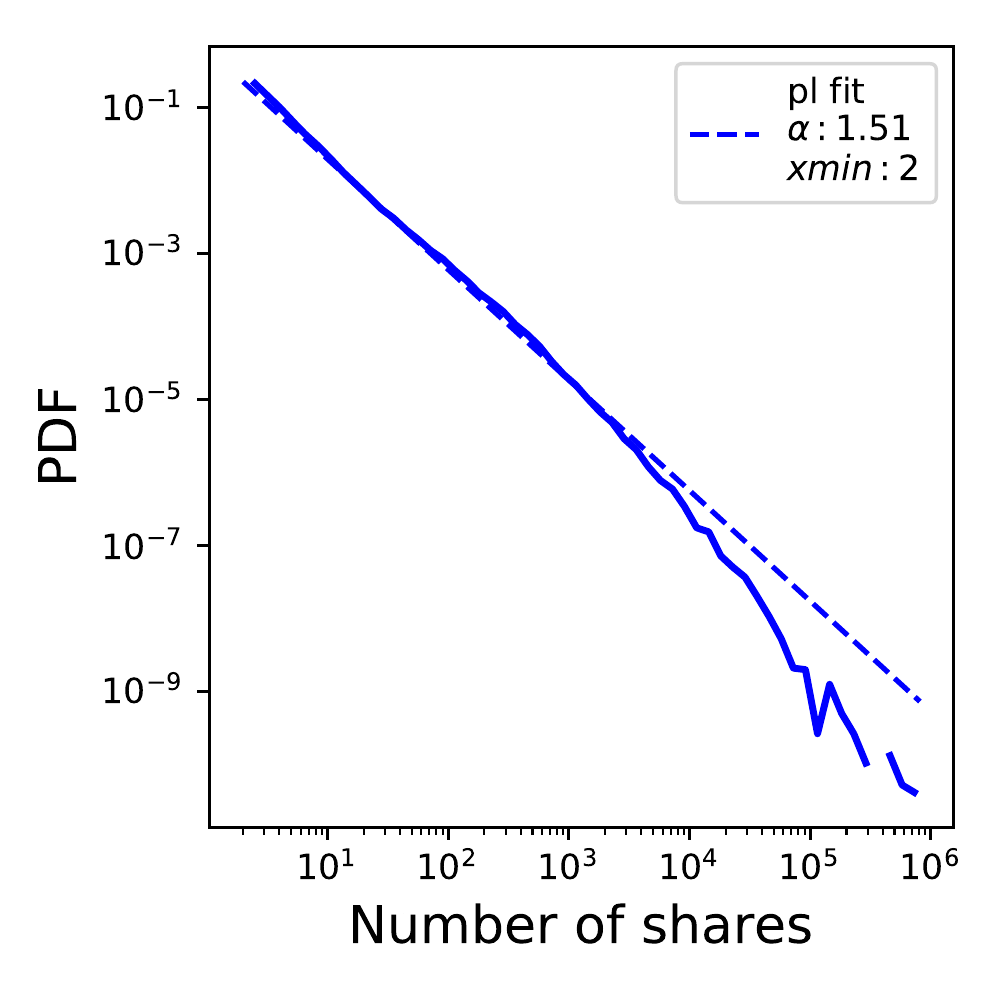}
   \caption{Facebook.}
   \label{fig:facebook} 
\end{subfigure}
\begin{subfigure}[b]{0.24\linewidth}
   \includegraphics[clip, trim=0cm 0cm 0cm 0cm,width=1\linewidth]{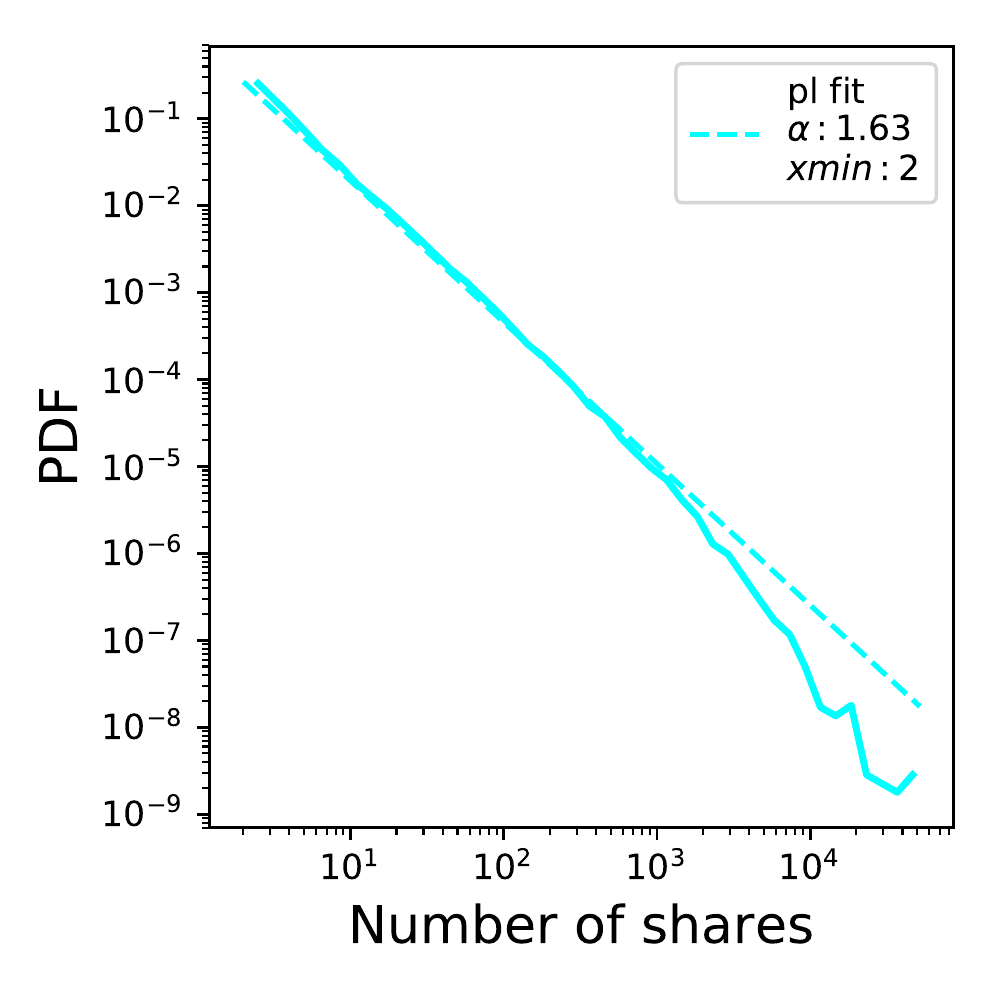}
   \caption{Twitter.}
   \label{fig:twitter} 
\end{subfigure}
\begin{subfigure}[b]{0.24\linewidth}
   \includegraphics[clip, trim=0cm 0cm 0cm 0cm,width=1\linewidth]{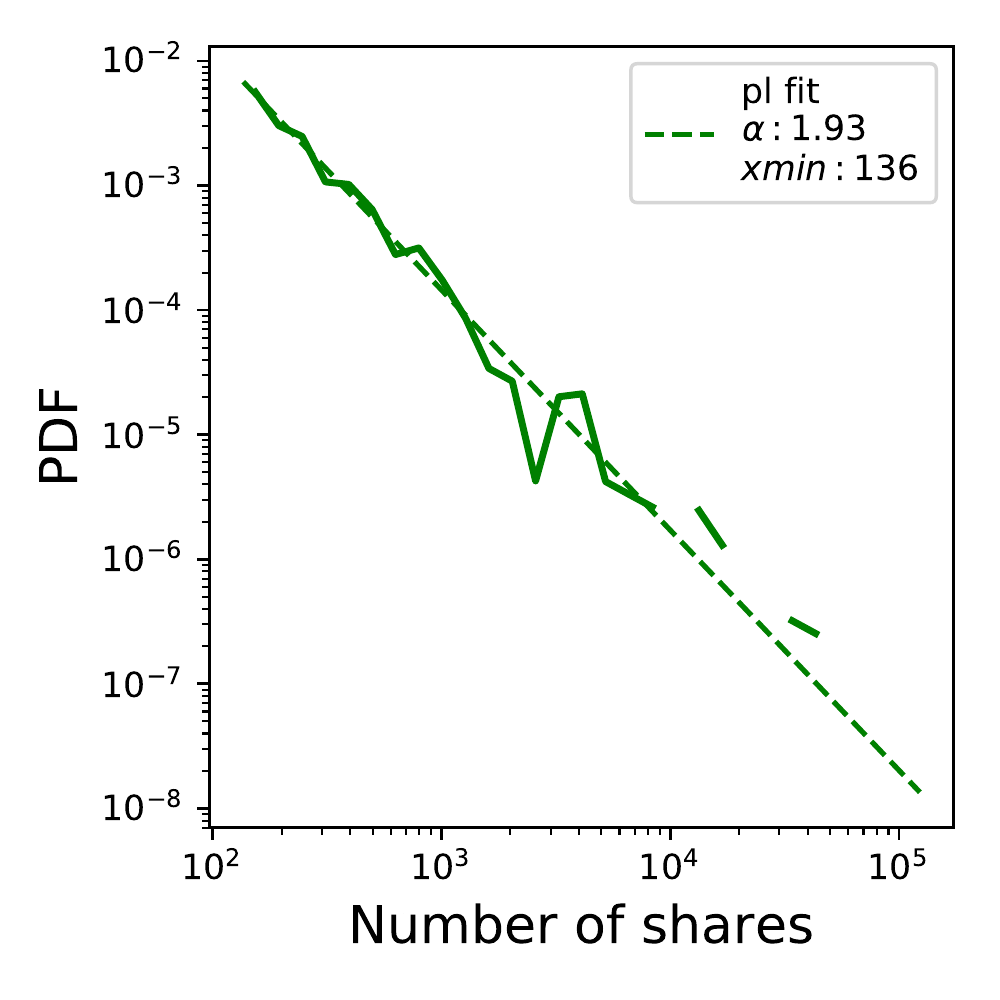}
   \caption{WhatsApp.}
   \label{fig:whatsapp} 
\end{subfigure}
\begin{subfigure}[b]{0.24\linewidth}
   \includegraphics[clip, trim=0cm 0cm 0cm 0cm,width=1\linewidth]{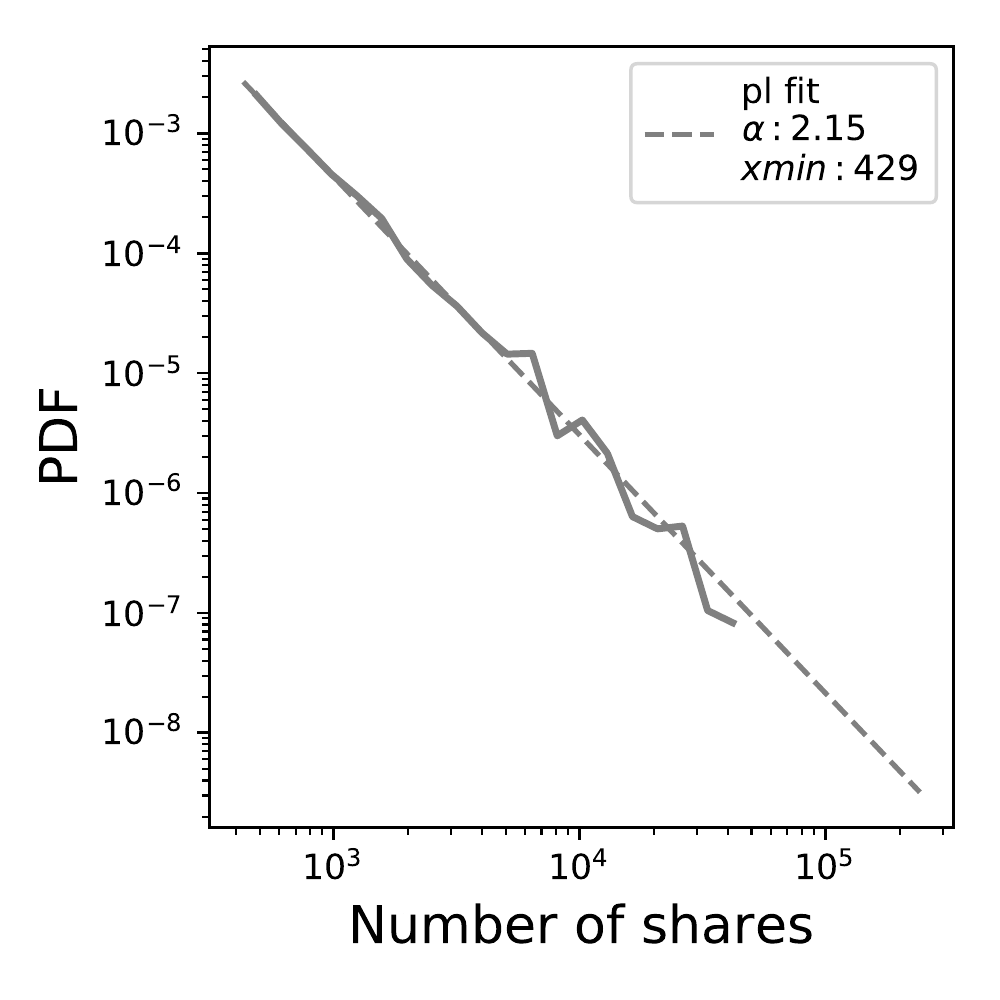}
   \caption{Email.}
   \label{fig:email} 
\end{subfigure}
\caption{Distribution of petitions by number of social media shares.}
\label{fig:shares}
\end{figure*}

\subsection{The link between social media campaigning and the success of online petitions}

Recent research has found that the growth and success of online petitions is influenced by their popularity in social media~\cite{margetts2015political,proskurnia2016please,Proskurnia:2017:PSO:3038912.3052705}. Given that our dataset includes how many times each petition was shared on Facebook, Twitter, WhatsApp and email, we present in Figure~\ref{fig:shares} the distribution of petitions by the number of shares in each platform. The plots show heavy-tailed distributions: most petitions are not shared on social platforms while a few are highly shared. This is consistent with previous studies of diffusion on social media~\cite{Goel:2012:SOD:2229012.2229058,Hale2018how}.

\begin{figure}[t]
\centering
\begin{subfigure}[b]{0.45\linewidth}
   \includegraphics[clip, trim=0cm 0cm 0cm 0cm,width=1\linewidth]{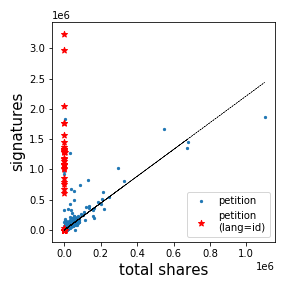}
   \caption{Full dataset.\\~\\~}
   \label{fig:scatter_sign_shares_0} 
\end{subfigure}
\begin{subfigure}[b]{0.45\linewidth}
   \includegraphics[clip, trim=0cm 0cm 0cm 0cm,width=1\linewidth]{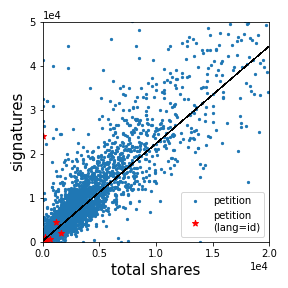}
   \caption{Subset (shares $\leq 2 \cdot 10^4$ and signatures $\leq 5\cdot 10^4$) covering 99.8\% of the full dataset.}
   \label{fig:scatter_sign_shares_1} 
\end{subfigure}
\caption{Signatures versus shares in social media. The black line is a linear fit, and petitions written in Indonesian are shown with red star markers.}
\label{fig:scatter_sign_shares}
\end{figure}

We explicitly examine the link between social media and the success of online petitions with a scatter plot of signatures versus the sum of shares on four social platforms (see Figure~\ref{fig:scatter_sign_shares_0}). Although the variables are positively correlated ($r=0.44, p < 0.001$), as better shown in Figure~\ref{fig:scatter_sign_shares_1}, we find of great interest the existence of a group of 25 petitions that received more than 500K signatures but less than 100 shares. We inspected these petitions and found that 24 of them were written in Indonesian by 18 authors not located in Indonesia but in the United Kingdom, the United States, Spain, France, Italy, Costa Rica and the Palestinian territories. Using the \texttt{author} field (given name and the initial of the family name), we searched Google using queries of the format: \emph{site:linkedin.com avaaz \texttt{name}}. For each query, we found an employee of \textit{Avaaz.org} (e.g., Campaign Directors, Senior Campaigners) matching with \texttt{name} and the geographical location of corresponding petitions. We provide two possible explanations for these results\footnote{After submitting the camera-ready copy of this article, the actual cause, described in the appendix, turned out to be neither of these two.}. On the one hand, they could be an indicator of astroturfing, i.e., these petitions could have been massively signed in an artificial manner. We should note that, in addition to the aforementioned problems of transparency and accountability~\cite{horstink2017online}, there are specific complaints about the reliability of the number of signatures  to petitions on \textit{Avaaz.org}~\cite{blogpost}. On the other hand, \textit{Avaaz.org} and other websites have been blocked in Indonesia in recent years~\cite{wiki:id}, and this finding could be the result of very effective campaigns through alternative (even non-digital) diffusion channels.

\begin{figure*}
\centering
\begin{subfigure}[b]{0.45\linewidth}
   \includegraphics[clip, trim=0cm 4cm 0cm 1.5cm,width=1\linewidth]{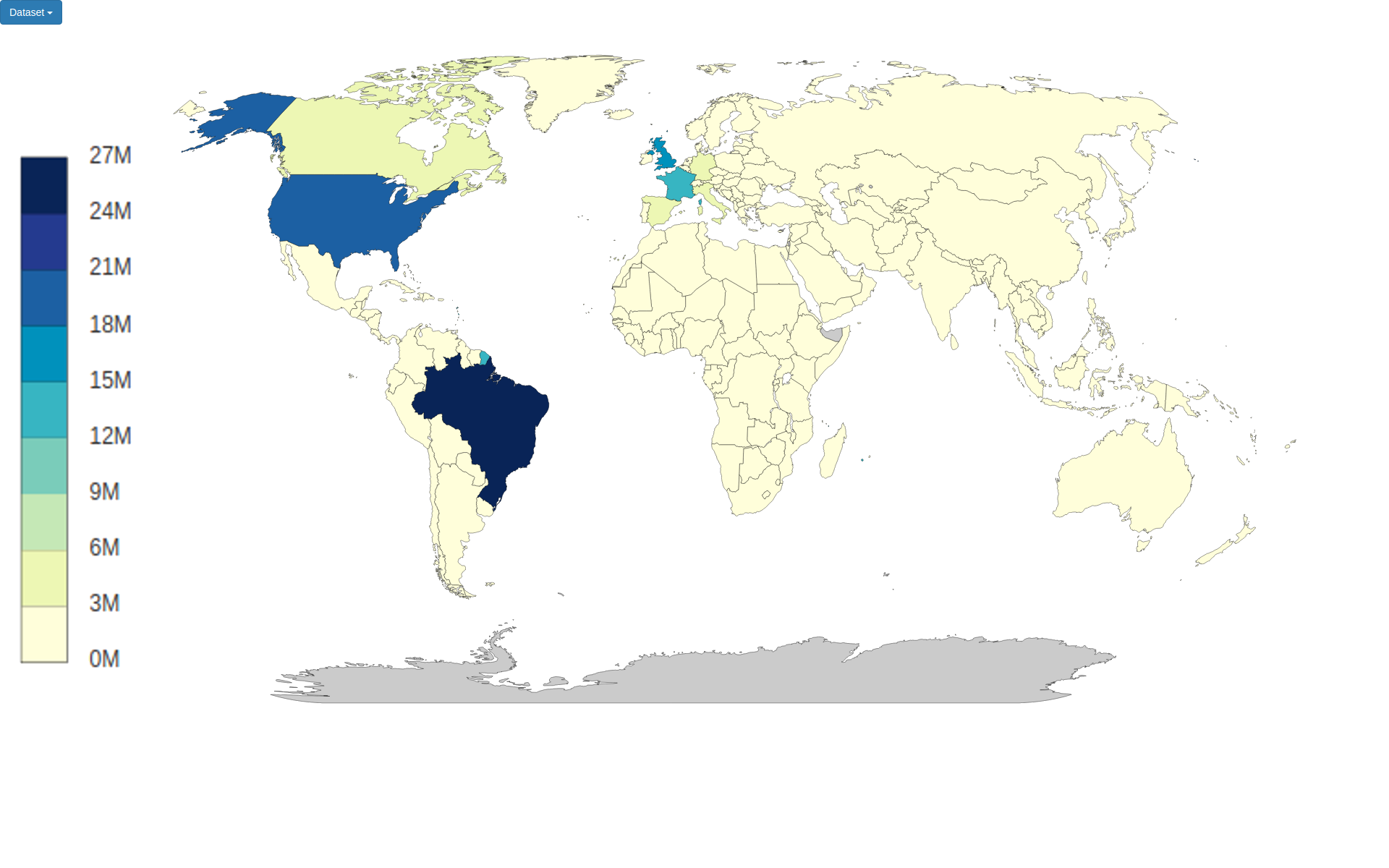}
   \caption{Sum of signatures (in millions).}
   \label{fig:map1} 
\end{subfigure}
\begin{subfigure}[b]{0.45\linewidth}
   \includegraphics[clip, trim=0cm 4cm 0cm 1.5cm,width=1\linewidth]{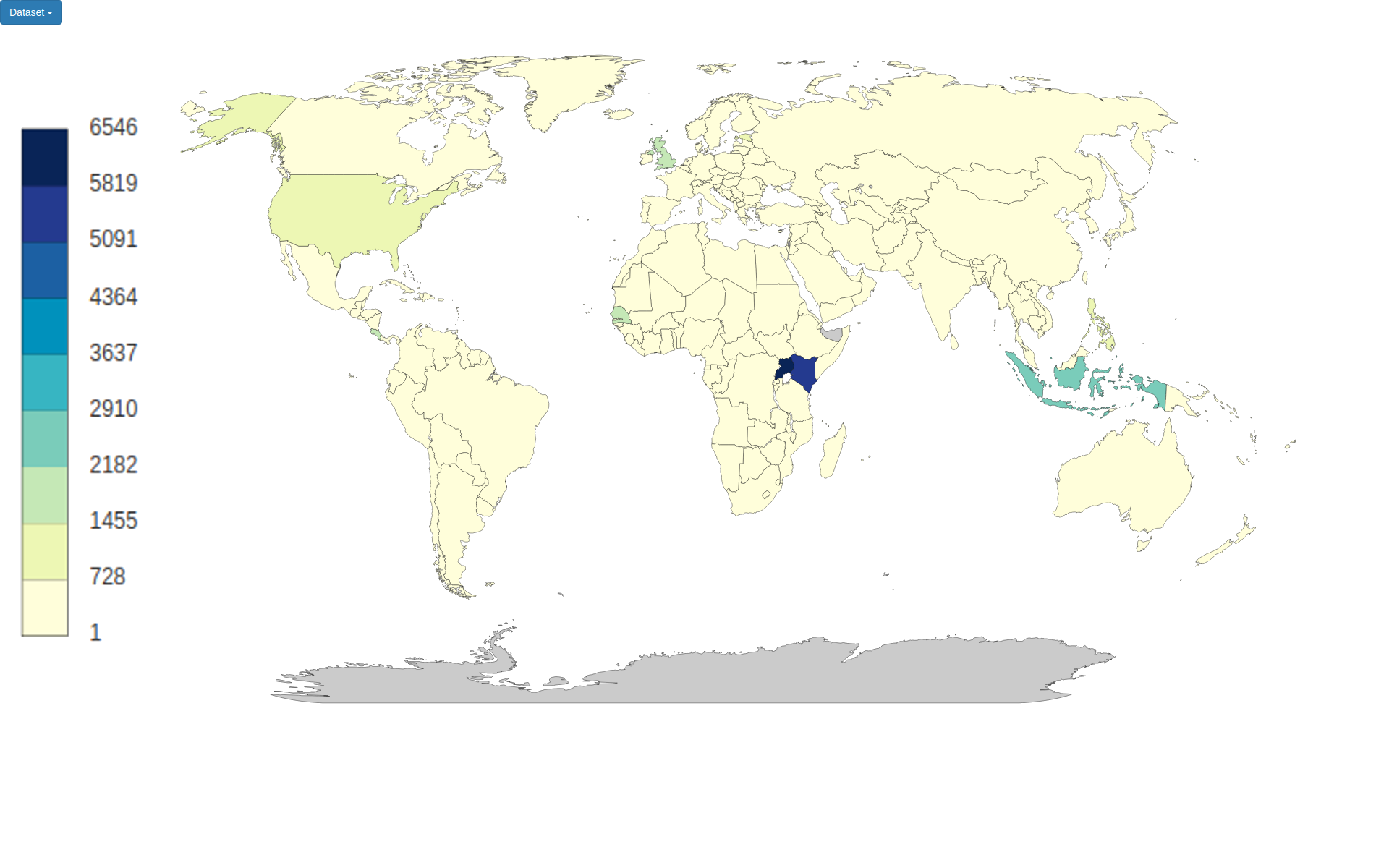}
   \caption{Average number of signatures per petition.}
   \label{fig:map2}
\end{subfigure}
\caption{Choropleth world maps of signing activity in \textit{Avaaz.org} by country.}
\label{fig:maps}
\end{figure*}

\subsection{Geographical and multilingual findings in a worldwide community}
Because the \textit{Avaaz.org} community is present in over 200 countries, we show in Figure~\ref{fig:maps} two choropleth world maps comparing activity across countries. The left map indicates the total number of signatures for all petitions in the dataset and reveals that activity is intense in Brazil, Spain, and countries that are members of the Group of Seven (G7) excluding Japan (Canada, United States, United Kingdom, France, Italy, and Germany). These results are similar to the map of the number of Avaaz `members' from each country in the Avaaz Annual 2016 Poll~\cite{avaaz2016}. In contrast, the right map shows the average number of signatures per petition and depicts a very different distribution with the highest values in two African countries, Uganda and Kenya, followed by Indonesia.

\begin{figure}[h]
 \centering
   \includegraphics[clip, trim=3.5cm 0cm 6cm 0cm, width=0.9\linewidth]{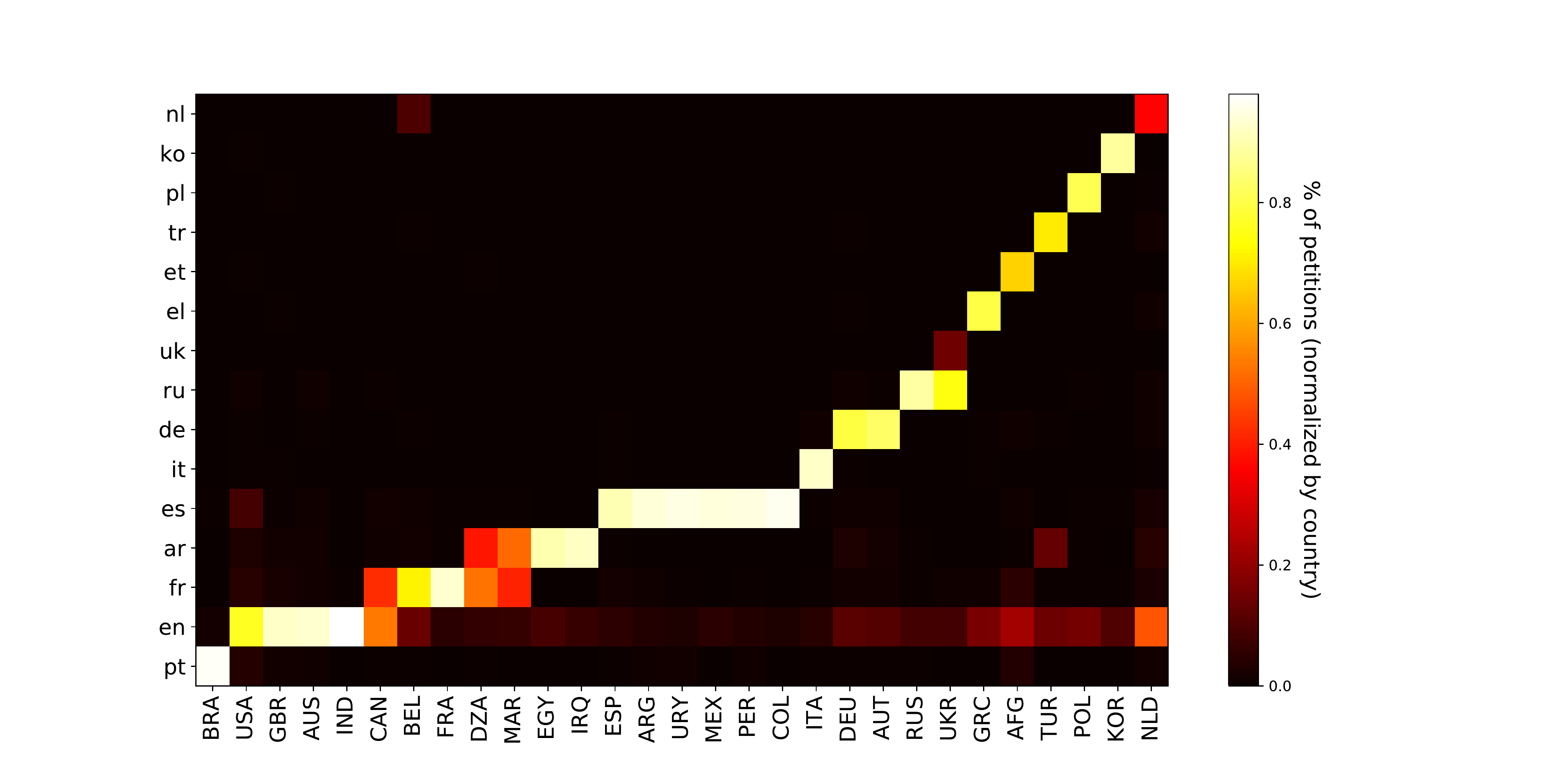}
 \caption{Heatmap of the percentage of petitions from different countries written in the most popular languages.}
\label{fig:heatmap_lang_country}
\end{figure}

To examine the most popular languages for \textit{Avaaz.org} petitions from very active countries, we present a heatmap in Figure~\ref{fig:heatmap_lang_country}.
For better readability, data are normalized by the number of petitions in each country, and countries on the horizontal axis are grouped by linguistic similarity. We observe that, although users tend to use the most spoken language in their countries, English acts as a global language with remarkable use in countries like Afghanistan, Poland, and the Netherlands. The results also allow for the easy identification of strongly multilingual countries, e.g., Canada (French and English), Argelia (French and Arab), Morocco (French and Arab), Ukrania (Russian and Ukranian), Belgium (French, Dutch and English) and Turkey (Turkish, Arab and English). 

\begin{figure}[t]
 \centering
   \includegraphics[clip, trim=1cm 0cm 3cm 1cm, width=0.9\linewidth]{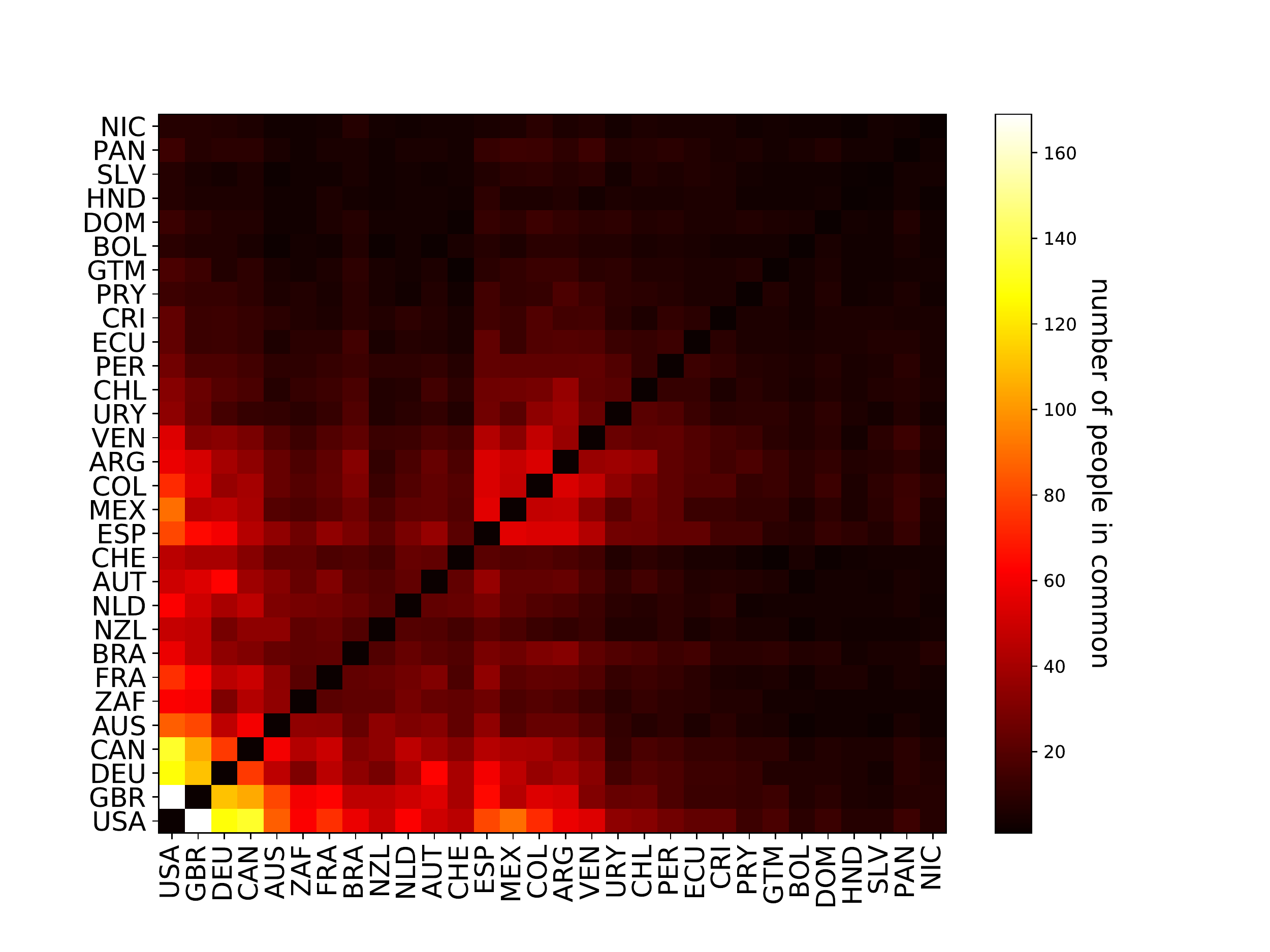}
 \caption{Heatmap of the number people who are named in petitions from multiple countries. The chosen countries have the largest number of petitions written in English, German or Spanish (the available languages of the Named Entity Recognizer). For better readability, values on the diagonal are set to 0.}
\label{fig:heatmap_people}
\end{figure}

Finally, we explore named people within the text of petitions. As indicated in the second section, this exploration is limited to petitions written in English, Spanish or German. Table~\ref{tab:people} shows the top 10 people who are mentioned in petitions from the largest number of countries. The table also includes the top 3 countries associated with each person. Because petitions in \textit{Avaaz.org} are aimed at solving global societal challenges, it was expected to find people associated to global leadership (e.g., Barack Obama or Angela Merkel). In addition to politicians, we find of interest the presence of Justin Bieber who appears in many petitions from Latin American countries like Argentina, Uruguay and Mexico. We inspect these petitions and found that many of them were about an indictment from an Argentinian court against him in 2016. This leads us to examine which countries share similar political and social references with a heatmap of the number of people in common between the most active English, Spanish and German speaking countries (see Figure~\ref{fig:heatmap_people}). Besides the overlap among countries very active on \textit{Avaaz.org} (United States, United Kingdom, Germany, Canada and Spain), we should note the specific overlap among Spanish speaking countries (Spain, Mexico, Colombia, Argentina, Venezuela, etc.) and between the United States and Mexico.

\begin{table}[h]
\centering
\caption{Top 10 people by the number of countries with English, Spanish or German petitions naming them. The last column indicates the three countries with more petitions naming the corresponding person (values in brackets).}
\label{tab:people}
\label{my-label}
\scalebox{0.75}{
\begin{tabular}{l|r|lr|lr|lr}
Person         & \begin{tabular}[c]{@{}c@{}}No. of \\ countries\end{tabular} & \multicolumn{6}{c}{Top 3 countries}  \\ \hline
\rule{0pt}{3ex}Barack Obama   & 94               & USA & (211) & DEU & (41) & GBR & (40) \\
Boris Johnson  & 71               & GBR & (60)  & USA & (15) & URY & (9)  \\
Angela Merkel  & 46               & DEU & (267) & AUT & (14) & GRC & (10) \\
Ban Ki-Moon    & 46               & USA & (12)  & GBR & (12) & CHE & (11) \\
David Cameron  & 41               & GBR & (471) & FRA & (7)  & USA & (5)  \\
Vladimir Putin & 41               & USA & (18)  & GBR & (14) & RUS & (11) \\
Edward Snowden & 31               & DEU & (69)  & USA & (18) & GBR & (14) \\
Justin Bieber  & 29               & ARG & (30)  & URY & (19) & MEX & (11) \\
Donald Trump   & 26               & USA & (17)  & GBR & (8)  & MEX & (6)  \\
Xi Jinping     & 26               & USA & (11)  & GBR & (5)  & CAN & (3) 
\end{tabular}
}
\end{table}

\section{Recommendations for Future Work}
In the above section we have presented a preliminary exploration of our dataset of online petitions from \textit{Avaaz.org}. On the basis of these results, we conclude this article by proposing different example research questions to reflect potential uses of the dataset.

\subsubsection{Bot detection} In the current context of the rise of social bots~\cite{Ferrara:2016:RSB:2963119.2818717}, we have provided the first evidence of bots on a petition platform by examining the most active users. Nevertheless, \textit{could bot-generated petitions also be detected with other (e.g., text-based) features?}. Indeed, our dataset is not only affected by bots publishing a large number of petitions but also by suspiciously high levels of support to petitions with almost no traction on social media. If astroturfing is the reason behind this pattern, and as done on other platforms like Twitter~\cite{Truthy_icwsm2011class}, \textit{could an automatic classifier identify petitions with artificial support?} Given the nature and potential of online petitions to influence policy makers, the answer to these questions would have clear political and social implications. 

\subsubsection{Content analysis} Our exploration of the content of online petitions was limited to detected languages and named entities. This has allowed us to obtain a geographic overview of the \textit{Avaaz.org} community, revealing patterns of multilingualism and common political references among countries. However, the textual content of the petitions still has great informative value to be examined. First, future work might focus on sentiment analysis. Given that recent research on \textit{Change.org} found that petitions were more likely to be successful when having positive emotions~\cite{7427430}, \emph{is this a specific finding of Change.org or also valid for the community of Avaaz.org?} 

Second, topic modeling could also be of great interest. For example, given the topics of online petitions from a multilingual country, \textit{do different linguistic communities care about the same issues?} Our dataset could help to identify socio-political problems of these communities. Furthermore, detected topics along with \texttt{date} field values would allow modeling the rise and decay of topics: \textit{how do topics emerge in online petition platforms?} According to \textit{Avaaz.org}, overall priorities for campaigns are set through member polls~\cite{avaaz2018about}. However, the motivations of members in selecting priority issues remain unclear. For the same period of time, a comparison of topics from our dataset to topics from external sources (e.g., news repositories, social media) would shed light on political agenda setting in the era of global activism. 

\subsubsection{Gender studies} Because \texttt{author} in our dataset is essentially the given name of the user who published a petition, future work might also focus on extracting the gender of authors to then identify whether there are significant differences in the topics about which men and women create petitions. A recent study on \textit{Change.org} found that women were more likely to publish petitions about animals and women's rights while men focused on economic justice and (general) human rights~\cite{peixoto}. Besides comparing whether these results are also valid in \textit{Avaaz.org}, the geographical and linguistic information of our dataset would allow for the investigation of a more detailed research question: \textit{is the distribution of relevant topics for men and for women stable over countries or affected by cultural factors?} Given the increasing awareness about gender biases online (e.g., Wikipedia~\cite{Wagner2016} and Facebook~\cite{garcia2017facebook}), our dataset might be a helpful resource to assess whether the democratic purpose of online petitioning is affected by any gender imbalances.

\section{Acknowledgment}
This work is supported by the Spanish Ministry of Economy and Competitiveness under the María de Maeztu Units of Excellence Programme (MDM-2015-0502), and the mobility grant offered by Societat Econòmica Barcelonesa d'Amics del País. We would also like to thank Chico Camargo for his valuable comments and suggestions.

\bibliographystyle{aaai}
\bibliography{references}

\section{Appendix: Comments provided by \textit{Avaaz.org}
about \textit{selenium s.} and the petitions with high signature counts but low social shares}

After the submission of the camera-ready copy, we were contacted by Ben Boyd (Chief Technology Officer at \textit{Avaaz.org}) who provided an explanation of the anomalies detected when exploring the dataset. 
First, some petitions authored by \textit{selenium s.} are likely the result of automated testing by \textit{Avaaz.org} which were not properly removed. However, bot activity is an ongoing challenge that \textit{Avaaz.org} is actively working to tackle. 
Second, the reason of the petitions with more than 500k signatures with extremely low social share counts is that \textit{Avaaz.org}'s campaigners linked some campaigns internally to the core campaigning platform to boost them. The linking procedure wrongly showed the number of signatures from the source petition of the campaign while not showing the corresponding share counts. This bug is now fixed (since June 2018). 
We would like to thank Ben Boyd for his insightful feedback and for his interest in this research project.
\end{document}